# Nanoscale Chemical Reaction Imaging at the Solid-Liquid Interface *via* TERS

*Ashish Bhattarai*[*] *and Patrick Z. El-Khoury*[*]

Physical Sciences Division, Pacific Northwest National Laboratory, P.O. Box 999, Richland, WA 99352, USA

[*]ashish.bhattarai@pnnl.gov; patrick.elkhoury@pnnl.gov

**ABSTRACT**

Plasmon-enhanced chemical transformations at the solid-liquid interface can be imaged with high sensitivity, chemical selectivity, and nanoscale precision through tip-enhanced Raman scattering (TERS). We demonstrate the latter for the first time through measurements aimed at (i) locating plasmonic hotspots at the solid-liquid interface at which chemical transformations take place, (ii) monitoring the evolution from reactants to products through their distinct Raman spectra, and (iii) 2D correlation analysis of Raman time trajectories to unambiguously extract the spectral components that mark chemical transformation and to understand the correlations between the product and parent signatures. For our proof-of-principle study, we select a model plasmon-enhanced chemical process, namely, the dimerization of *p*-nitrothiophenol to dimercaptoazobenzene, but now, at the solid-liquid interface. Our plasmonic construct otherwise consists of chemically functionalized gold microplates in aqueous solution, which we image using a gold-coated TERS probe irradiated at 633-nm. Overall, we demonstrate chemical reaction imaging in aqueous solution *via* TERS for the first time, herein, at a pixel-limited lateral spatial resolution of 10 nm.

**Introduction**

The powerful combination of scanning probe microscopy and Raman spectroscopy in the so-called tip-enhanced Raman scattering (TERS[1-4]) scheme has revolutionized chemical imaging. Sub-nm spatial resolution has been demonstrated using scanning tunneling microscopy-based TERS in ultra-high vacuum and at cryogenic temperatures.[5-8] Under ambient laboratory conditions, atomic force microscopy (AFM)-based TERS with single/few nanometer precision has also been reported,[9-11] which paves the way for broader applicability of this powerful technique. While chemical fingerprinting with high sensitivity and nanometer spatial resolution has been one of the drives behind advancements in TERS, chemical reaction imaging remains the ultimate goal. This is particularly the case in solution where most chemical transformations of interest to the biological and energy sciences take place. To date, the latter has not been demonstrated.

Several pioneering works illustrated that surface plasmon-enhanced Raman scattering can be used to monitor chemical transformations. Most recently, the plasmon-driven polymerization of thiols was visualized through TERS.[12] Other prior demonstrations include tracking tautomerization of porphycene at plasmonic hotspots,[13] plasmon-enhanced N-demethylation of methylene blue,[14] photo-induced and plasmon-driven *cis-trans* isomerization,[15,16] and photocatalytic reduction of $CO_2$ following visible light irradiation.[17] The oxidative and reductive dimerization of *p*-nitrothiophenol (NTP) and *p*-aminothiophenol (ATP) to dimercaptoazobenzene (DMAB) on plasmonic substrates has been extensively scrutinized by several groups.[18-21] Although controversial in its early days,[22-25] subsequent analyses of these reactions paths convincingly established plasmon-enhanced DMAB formation. Overall, the DMAB product is routinely observed throughout the course of plasmon-enhanced Raman scattering measurements targeting both ATP and NTP.[18-21,24,26] Several mechanistic investigations have subsequently been performed, with an emerging consensus that hot-electrons drive azo bond formation to yield DMAB.[21,27-30] Note that all of the aforementioned works investigated DMAB formation through (nominally diffraction-limited) surface-enhanced Raman scattering measurements that nonetheless track plasmon-induced chemical transformations with high sensitivity and chemical selectivity.

The first attempt at imaging the NTP → DMAB rection *via* TERS was reported by Lantman et al.[31] The prior investigation employed a combination of a 633 nm laser source to probe the parent (NTP) through TERS and a 532 nm light source to induce the dimerization process. More generally, TERS maps of catalytic hot-spots were recently recorded with a lateral spatial resolution of 20 nm.[32] Nonetheless, similar measurements in solution/at the solid-liquid interface continue to be scarce, particularly in the TERS scheme.[33-35] This is the subject of our present work, whereby we use *in situ* TERS nanoscopy[36] to both locate hotspots at which the NTP → DMAB reaction occurs and to examine the interplay between parent and product molecules – all at the solid-liquid interface.

**Results**

Figure 1 highlights our experimental platform. Figure 1A shows an AFM image of gold microplates supported by a glass coverslip immersed in $H_2O$. The gold microplate is imaged using a ~100 nm Au-coated AFM probe. The inset of Figure 1A shows a topographic cross-section taken towards the left edge of the plate. The cut reveals two stacked/displaced microplates, each of which is ~30 nm thick. Using an inverted optical microscopy scheme (bottom excitation-collection) with an in-plane polarized 633-nm laser source, we record hyperspectral TERS images of the rectangular area that is highlighted in Figure 1A. TERS results are shown in Figures 1B and 1C. The two panels shown simultaneously recorded TERS image slices at frequency shifts that can be assigned to the parent NTP (1330 cm$^{-1}$, Figure 1B) and product DMAB species (1425 cm$^{-1}$, Figure 1C).[21] Unless otherwise stated, our work is solely concerned with the *trans*-form of DMAB.

As a result of the inverted excitation-collection geometry, only the edges and their immediate vicinity are visible in our TERS maps.[36] With that in mind, the overall TERS image of the reactants (Figure 1B) traces the local optical fields that are sustained towards the edges.[36] This observation is consistent with several prior works from our group.[37,38] On the other hand, the simultaneously recorded TERS image at 1425 cm$^{-1}$ shows that product formation is favored at specifc sites on the substrate. In other words, the product map does not necessarily follow the overall spatial profile and magnitude of the local optical field. Spatio-spectral analysis in Figure 1D shows three consecutively recorded TERS spectra and further illustrate the concept. These spectra were extracted from three adjacent pixels (10 nm lateral step size) and clearly show pixel-to-pixel variation. These observations allow us to infer that the dimerization process is (i) site-specific, and (ii) generally occurs at some (not all) sites where electric fields are optimally enhanced. The latter can be rationalized by recognizing that the local environments at which chemical transformations take place are distinct. Moreover it appears that our images track the dimerization process at the solid-liquid interface with sub-10 nm resolution. Further evidence in support of this statement is given in the supporting information section, see Figure S2.

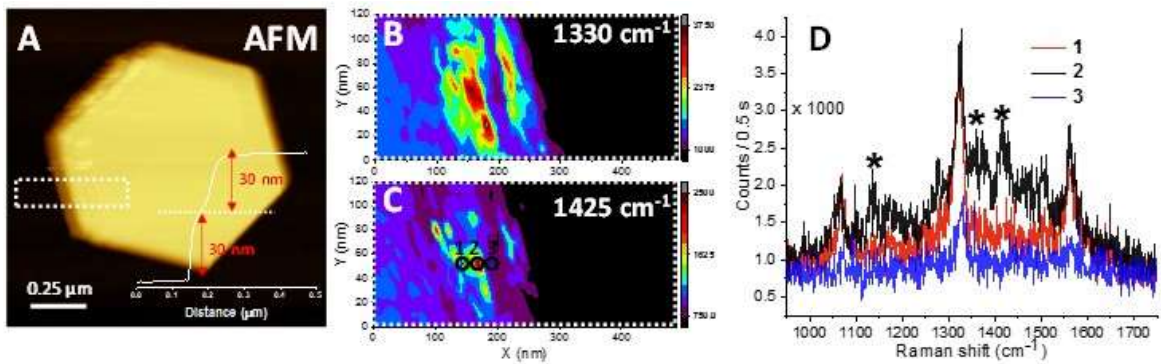

**Figure 1:** A) A representative AFM image of gold micro-plates in H$_2$O. The inset shows a topographic cross-sectional cut. The highlighted rectangular area in A) is visualized through TERS. Panels B) and C) show simultaneously recorded TERS images at 1330 and 1425 cm$^{-1}$. Panel (D) shows three spectra taken at adjacent pixels (marked by 1, 2, and 3 in C). The black asterisks (*) highlight the major Raman modes for DMAB. The TERS images were recorded with a lateral step size of 10 nm, and the spectra were integrated for 0.5 s at each pixel.

After a reaction cite is located (e.g., site-2 in Figure 1C), the tip is positioned at the identified position and brought into contact. Subsequently, sequential TERS spectra are recorded to track sluggish (limited by the integration time used) product formation kinetics. Figure 2A shows a waterfall plot of the Raman trajectory at site-2 (see Figure 1C), henceforth termed junction 1. In this case, the first 8 spectra/frames show a persistent presence of the parent (NTP), and no observable product (DMAB) peaks. Conversely, frames 9 through 14 show a persistent and dominant presence of DMAB bands for a time period of 2.5 seconds. After a brief disappearance of the product peak (frame 15), the DMAB signature is recovered at frame 16 and persists for a time period of 5 seconds. Throughout the remainder of this time trajectory, the product signal disappears, exposing the Raman signature of NTP. Figure 2B shows time-averaged spectra of NTP (first 8 frames) and DMAB (frames 9-14), which can be clearly delineated through their distinct Raman lines, *vide infra*.

The time evolving signals shown in Figure 2 suggest that the spectra of both parents and product arise from small ensembles. A more in-depth analysis of the DMAB spectra reveals that that the relative intensities of the 1130, 1370, and 1425 cm$^{-1}$ bands remain constant throughout the time trajectory, which supports our inference of an ensemble averaged optical response, see Figure S3. A cross-correlation slice at 1130 cm$^{-1}$ (a DMAB resonance) taken from a 2D correlation of the recorded time trajectory (see supporting Figure S4) is shown in Figure 2C. The trace shows that the product peaks are non-correlated to the parent peaks, e.g., the predominant 1330 cm$^{-1}$ NTP band. It therefore appears that a small fraction of NTP molecules in our probing volume have undergone dimerization into DMAB.

On the basis of the results shown in Figure 2, the fate of the product species can be understood in two ways. First, the conversion of NTP to DMAB may be reversible. Second, the product species can diffuse out of the effective probing area over the timescale of our measurement. The absence of correlation between the product and parent peaks, albeit suggestive of diffusion, cannot totally rule out the decomposition of DMAB into two NTP molecules at the junction. While some prior evidence supports DMAB photo-reduction into ATP at plasmonic junctions,[39,40] and to NTP in basic media,[41] we (i) do not observe the optical signatures of ATP, and (ii) perform our measurements in Millipore H$_2$O (Millipore, pH = 7). That said, our first trajectory cannot be used to rigorously preclude any of the mechanisms for product disappearance.

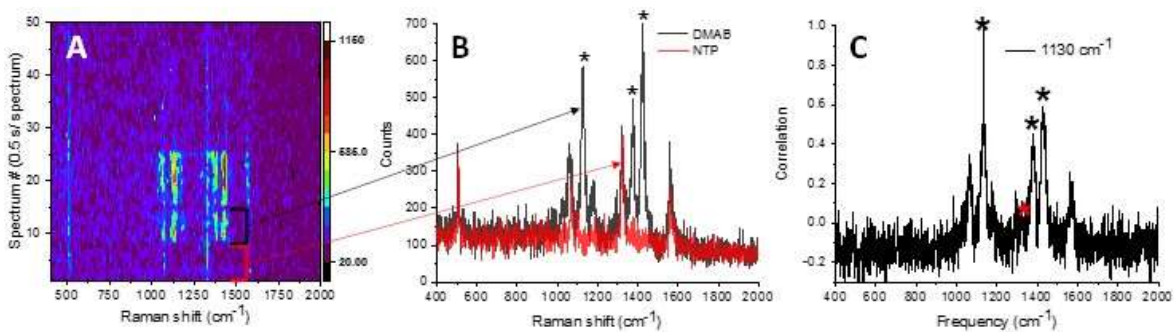

**Figure 2:** Panel (A) shows the time evolution of point TERS spectra at a hot-spot. Panel (B) shows two time-averaged TERS spectra of NTP (black) and DMAB (red). Panel (C) shows a cross-correlation slice at 1130 cm$^{-1}$ of the data presented in (A). Black asterisks denote the Raman modes

that are associated with the product, DMAB, whereas red asterisks mark the 1330 cm$^{-1}$ resonance of NTP. Each spectrum was integrated for a time period of 0.5 s.

Raman trajectories at other hotspots (e.g., junction 2) paint a different picture, see Figure 3. Unlike junction 1, sporadic DMAB formation events are observed herein. Three consecutive spectra shown in Figure 3B taken at 90$^{th}$, 91$^{st}$, and 92$^{nd}$ frames (0.25 s/frame in this case) illustrate the concept. On the basis of this data, it appears that all NTP molecules in our effective probing volume undergo dimerization. The observation of similar patterns in a previous SERS study[21] was associated with single/few molecular events at plasmonic hotspots. Unlike the case at junction 1, the relative intensities of the Raman lines of the DMAB product exhibit pronounced intensity fluctuations (see Figure S5), which was also previously associated with Raman scattering in single/few molecule regime.[42] A 2D correlation analysis of our Raman time trajectory aids in the interpretation of temporally varying parent and product signatures, see Figures 3C and 3D. A cross-correlation slice at 1130 cm$^{-1}$ (a DMAB resonance) shows that the product peaks at 1130, 1370, and 1425 cm$^{-1}$ are strongly correlated to each other. This is expected, as the lines arise from the same species. Interestingly, the same slice shows an inverse correlation between the product and parent peaks (e.g., the predominant 1330 cm$^{-1}$ NTP resonance). These results paint a different picture than the one for junction 1, see above. Namely, DMAB formation at junction 2 depletes the parent NTP species therein.

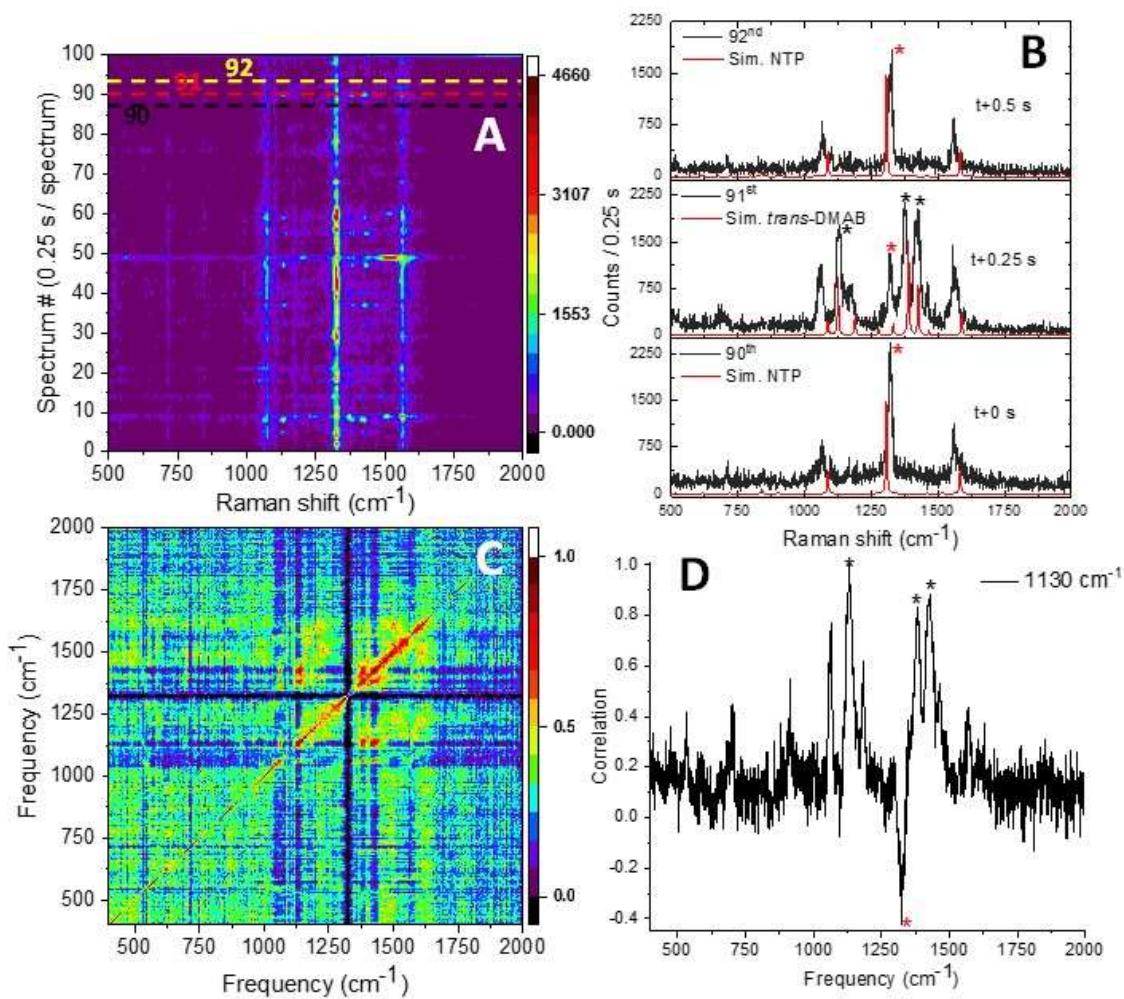

**Figure 3:** Panel A shows a TERS spectral trajectory at a hot-spot (junction 2). Panel B shows sequentially recorded TERS spectra (black) at 90$^{th}$, 91$^{st}$, and 92$^{nd}$ time series. The red spectra in this panel correspond to PBE/def2-TZVP Raman spectra of the parent and product species. A 2D correlation analysis of the time trajectory is shown in panel C and a cross-correlation cut at 1130 cm$^{-1}$ is shown in panel D. Black and red asterisks highlight the peaks associated with the product and parent, respectively. Each spectrum was integrated for a time period of 0.25 s.

Throughout the course of our TERS measurements, we additionally occasionally observed a unique set of vibrational lines at 1465 cm$^{-1}$ and 1500 cm$^{-1}$ (Figure 4C, red spectrum) that cannot be assigned to NTP or DMAB. Note that up to this stage, all of our discussion and assignments were concerned with *trans*-DMAB. While it is not uncommon for azobenzenes to undergo *cis-trans* isomerization,[16,43] the majority of the reported plasmon-enhanced NTP → DMAB conversion processes in literature implicate *trans*-isomer of the product, besides a few theoretical demonstrations.[44,45] Namely, potential energy profiles obtained upon rotation around the N=N bond showed that the *cis-* isomer of an azobenzene is less stable than its *trans-* form.[46] Figure 4A and B show simultaneously recorded TERS maps at 1330 cm$^{-1}$ and 1500 cm$^{-1}$ around the edge of a gold microplate. Much like in Figure 1, we show two simultaneously recorded maps at 1330 and 1500 cm$^{-1}$. Figure 4C shows 4 spectra representing experimental NTP (position 1 in Figure 4A and B) and product (position 2 in Figure 4A and B) spectra, along with their theoretical analogues. Although the relative intensities of the predicted peaks for the *cis*-DMAB species are not in agreement with their theoretical analogues, the measured (1465 cm$^{-1}$ and 1500 cm$^{-1}$) and predicted vibrational resonances (1455 cm$^{-1}$ and 1488 cm$^{-1}$) are reasonably well-aligned. We tentatively assign the product spectrum to *cis*-DMAB on this basis. Difference in the relative intensities of the observable states can otherwise be rationalized on the basis of modified TERS selection rules[38] and/or the removal of orientational averaging in our case of Raman scattering from a nanoscopic probing volume.[47]

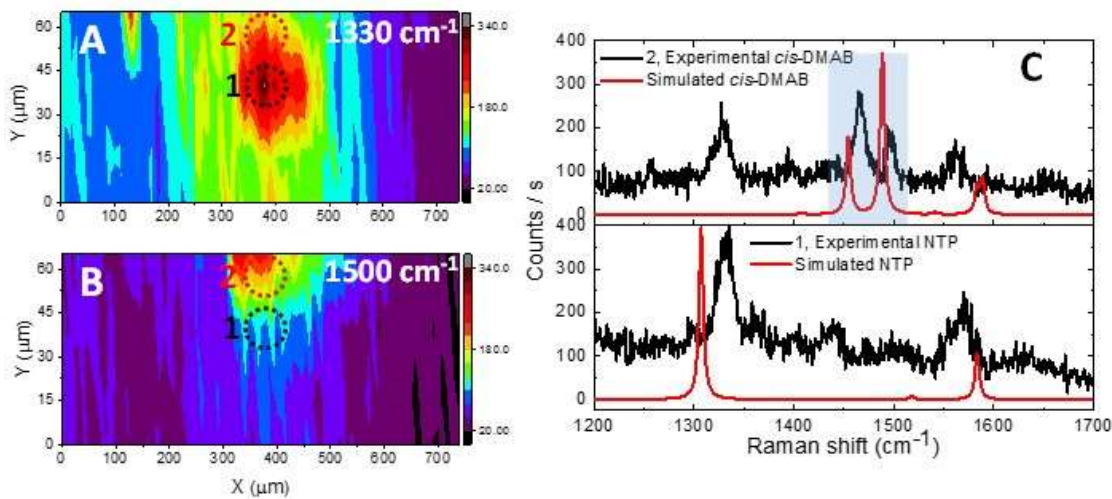

**Figure 4:** Simultaneously recorded hyperspectral TERS images of an edge of a gold microplate at 1330 cm$^{-1}$ (A) and 1500 cm$^{-1}$ (B). Panel (C) shows TERS spectra (black) at positions 1 and 2, as denoted in (A) and (B), along with the PBE/def2-TZVP Raman spectra of the two species (red). The highlighted area in panel (C) isolated the peaks of *cis-* DMAB. Each spectrum was integrated for a time period of 1 s.

**Conclusions**

We describe a platform and protocol that may be used to image chemical reactions at the solid liquid interface *via* TERS. Our work illustrates how plasmonic hotspots at which chemical transformations take place can be located through *in situ* TERS imaging. Subsequently, the tip can be brought into contact at the site(s) of interest to track the time evolution of product formation at a plasmonic nanojunction. We observe trajectories suggestive of both ensemble-averaged and single/few molecule dimerization events, tracked through TERS trajectories and correlation analysis. Overall, our results show that *in situ* TERS at the solid-liquid interface may be used for more than mere chemical identification and imaging. Chemical transformations can be tracked through TERS in solution under ambient laboratory conditions.

**Methods**

Gold nanoplates (AP1-10/1000-CTAB-DIH-1-5, NanoPartz$^{TM}$) stock is deposited and allowed to dry onto a glass bottom solution cell (Cellvis), followed by sonicating the cell in ethanol for ~30 seconds and rigorously washing the substrate using the same solvent. Subsequently, 100 uL of 3 mM ethanolic 4-nitrothiophpenol (NTP, Tokyo Chemicals Industries) solution was added and allowed to react for an hour. The mixture is then sonicated for ~30 seconds and washed with ethanol. The alcohol is allowed to dry prior to adding 150 µl of Millipore water for the liquid TERS measurements that are described below.

Our AFM/TERS setup is described elsewhere.[10,37,38] For the purpose of this work, initial topographic AFM measurements were performed in tapping mode feedback using a silicon tip (Opus, 160AC-NN) coated with 100 nm of Au by arc-discharge physical vapor deposition (target: Ted Pella Inc., 99.99% purity). A 633 nm laser (100-200 µW) is incident onto the apex of the TERS probe using a 100X air objective (Nikon, NA=0.85) using the bottom (inverted) excitation channel. The polarization of the laser is controlled with a half waveplate and is orthogonal to the long axis of the AFM probe (in-plane). The scattered radiation is collected through the same objective and filtered through a series of filters. The resulting light is detected by a CCD camera (Andor, Newton EMCCD) coupled to a spectrometer (Andor, Shamrock 500). A dedicated TERS imaging mode (SpecTop, patent pending from AIST-NT) was employed for simultaneous AFM-TERS mapping in $H_2O$. Using this mode, TERS signals are collected when the tip is in direct contact with the surface with a typical force in the 10−25 nN range. A semicontact mode is then used to move the sample relative to the tip (pixel to pixel) to preserve the sharpness and optical properties of the tip and to minimize the lateral forces that otherwise perturb the substrate.

**Author Contributions.** PZE and AB jointly designed and performed the experiments, analyzed the data, and wrote the manuscript.

**Acknowledgments.** AB and PZE were supported by the Department of Energy's (DOE) Office of Biological and Environmental Research Bioimaging Technology project #69212. Part of the instrumentation that is required to perform TERS in solution was purchased through support from the PNNL LDRD program. This work was performed in the environmental and molecular sciences laboratory (EMSL), a DOE Office of Science User Facility sponsored by BER and located at PNNL. PNNL is operated by Battelle Memorial Institute for the DOE under contract number DE-AC05-76RL1830.

**Competing Interests**. The authors declare no competing financial interests.

**Correspondence**. Correspondence and requests for materials should be addressed to Patrick Z. El-Khoury (Patrick.elkhoury@pnnl.gov) or Ashish Bhattarai (ashish.bhattarai@pnnl.gov).

**Data Availability.** Data presented in this work is available upon request from Patrick Z. El-Khoury or Ashish Bhattarai.